\documentclass[aps,prb,showpacs,
floatfix,
amsmath,amssymb,
               reprint]{revtex4-1}
\usepackage{bbold}
\usepackage{color}
\usepackage{graphicx}
\usepackage{natbib}
\usepackage{setspace}
\usepackage{amsmath}
\usepackage{amssymb}
\usepackage{verbatim}
\usepackage{bibentry}

\def\nnu{{\nonumber}}
\def\bk{{\mathbf{k}}}
\def\bK{{\mathbf{K}}}
\def\bkt{{\mathbf{\tilde{k}}}}
\def\bR{{\mathbf{R}}}
\definecolor{scarred}{rgb}{0.75,0.0,0.0}

\begin{document}


\title{Localization of phonons in mass disordered alloys- A typical medium dynamical cluster approach}
\author{Wasim Raja Mondal}
\affiliation{Jawaharlal Nehru Centre for Advanced Scientific Research, Bangalore 560 064, India.}
\author{ T.\ Berlijn} 
\affiliation{Center for Nanophase Materials Sciences, Oak Ridge National Laboratory, Oak Ridge, Tennessee 37831}
\affiliation{Computational Sciences and Engineering Division, Oak Ridge National Laboratory, Oak Ridge, Tennessee 37831, USA}
\author{Juana Moreno}
\affiliation{Department of Physics \& Astronomy, Louisiana State University, Baton Rouge, Louisiana 70803}
\affiliation{Center for Computation \& Technology, Louisiana State University, Baton Rouge, Louisiana 70803, USA}
\author{M.\ Jarrell} 
\affiliation{Department of Physics \& Astronomy, Louisiana State University, Baton Rouge, Louisiana 70803}
\affiliation{Center for Computation \& Technology, Louisiana State University, Baton Rouge, Louisiana 70803, USA}
\author{N.\ S.\ Vidhyadhiraja} \email{raja@jncasr.ac.in}
\affiliation{Jawaharlal Nehru Centre for Advanced Scientific Research, Bangalore 560 064, India.}

\begin{abstract}
The effect of disorder on lattice vibrational modes has been a topic of interest for several decades. In this work, we employ a Green's function based approach, namely the dynamical cluster approximation (DCA), to investigate phonons in mass disordered systems. Detailed benchmarks with previous exact calculations are used to validate the method in a wide parameter space. An extension of the method, namely the typical medium DCA (TMDCA), is used to study Anderson localization of phonons in three dimensions. We show that, for binary isotopic disorder, lighter impurities induce localized modes beyond the bandwidth of the host system, while heavier impurities lead to a partial localization of the low frequency acoustic modes. For a uniform (box) distribution of masses, the physical spectrum is shown to develop long tails comprising mostly localized modes. The mobility edge separating extended and localized modes, obtained through the TMDCA, agrees well with results from the transfer matrix method.  A re-entrance behavior of the mobility edge with increasing disorder is found that is similar to, but somewhat more pronounced than, the behavior in disordered electronic systems. Our work establishes a new computational approach, which recovers the thermodynamic limit, is versatile and computationally inexpensive, to investigate lattice vibrations in disordered lattice systems.
\end{abstract}
\maketitle



\section{Introduction}
Anderson localization (AL) \cite {PhysRev.109.1492}, though over five decades old, generates sustained interest due to its importance in diverse phenomena such as metal-insulator transitions, quantum Hall effect, mesoscopic fluctuations in small conductors and quantum chaos. Being a wave phenomenon in disordered systems, AL is, naturally, not limited to electronic systems and has been found in many other systems like electromagnetic waves \cite{Nature446.52.55,Nature404.850.853}, acoustic waves \cite {Nature4.945.948}, and spin waves \cite{PhysRevB.33.6541}. Thus, the physics of AL is of direct relevance to various applications such as optical fiber design \cite {Opt.Express}, molecular spintronics and even in biological systems 
  \cite {CommunTheorPhys.61.755}.

A theoretical understanding of AL remains a challenging research topic though it has been pursued extensively over the years.  In this context, several computational techniques including exact diagonalization (ED), transfer matrix method, kernel polynomial method, \cite{Rep.prog,Intermod,PhysRevE.56.4822,goossens93,
PhysRevLett.47.1546,Kramer,mack,slev,pmark,slevin} and renormalization group method \cite{PhysRevB.20.4726,PhysRevB.33.7738,Asingh} have been developed and applied. A majority of these studies deal, however, with electronic systems, and less attention has been paid to other relevant elementary excitations, such as phonons, despite being accessible to experiment and having various applications like in high-performance thermoelectric materials design.

The present work aims to apply a recently developed framework, namely the typical medium dynamical cluster approximation (TMDCA), to investigate the AL of phonons in mass disordered alloys. In this section, we briefly introduce the problem and review the relevant work on phonon localization before delving into the formalism in the next section.

A random substitution of ions in a crystal lattice creates local disturbances, the extent of which depends on both the size and chemical nature of the impurity ions.
As a result, real space perturbations of a given unit cell can propagate to neighboring unit cells and be extended over a characteristic length scale $\xi$. If this length scale is comparable to the system size, the normal modes of the disordered system are termed extended, and adiabatic continuity can be expected to connect the disordered system with the clean case. However, it may happen that, at and beyond some critical value of the disorder strength, some
or all of these modes remain confined over a finite localization length, implying a real space localization of such modes.

This kind of disorder-induced confinement of lattice waves indicates localization of phonons. If impurities are heavier than host atoms, the phonon spectrum will be, in general, shifted towards low-frequency regions.  Lighter impurities, on the other hand, can lead to more interesting effects. New states corresponding to the vibration of guest atoms can appear in frequency regions where no levels of the host crystal were present. Hence new impurity bands isolated from the host-dominated spectra may be observed in the phonon spectrum. Thus, a small amount of disorder in lattice vibrations can change the physical properties of the material. For example, the introduction of impurities can dramatically reduce the thermal conductivity \cite{PhysRevB.91.094307,PhysRevB.61.3091,:/content/aip/journal/jap/107/5/10.1063/1.3329541,doi:10.1021/nl101836z, nat}, which is a key factor in the design of high-performance thermoelectric materials. 

Several experimental studies have been devoted to understanding the
AL of phonons and the effect of isotopic disorder \cite{PhysRevLett.14.633,PhysRevLett.70.1715,PhysRevB.44.12046,PhysRevB.53.13672}. Recently, Howie et al. \cite {PhysRevLett.113.175501} found direct experimental evidence of phonon localization in a dense Hydrogen-deuterium binary alloy.
Sarpkaya et al. \cite{doi:10.1021/acsnano.5b01997} observe that wave functions corresponding to acoustic phonons are strongly spatially localized 
in copolymer-wrapped carbon nanotubes. The first observation of localization of sound was made by Hu et al in a random three dimensional elastic network \cite{natsound}.
Very recently, Mendoza et al. \cite{doi:10.1021/acs.nanolett.6b03550} observed a strong effect of the AL of phonons on the thermal conductivity in GaAs/AlAs superlattices. 
A low temperature plateau in the thermal conductivity of disordered materials such as glasses \cite{PhysRevB.34.5696} and high-temperature superconductors \cite{PhysRevB.67.857} has been attributed to the AL of phonons. These experimental observations have not yet received a comprehensive theoretical treatment.  
 
Extensive theoretical attempts to investigate isotopic disorder exist and some even predate Anderson's work on localization. Most of these may be classified as either Green's function based approaches or computational methods. The former include perturbative, semi-analytical approaches\cite{PhysRev.92.1331,PhysRev.101.19,Maradudin1966273,PhysRev.100.525,WEISS1958327,taylor,PhysRev.131.163,RevModPhys.23.287,PhysRev.156.1017,PhysRevB.38.4906,PhysRevB.10.1190,0022-3719-6-10-003,0022-3719-17-6-010,PhysRevB.21.4230,PhysRevB.18.5291,PhysRevB.66.214206} and continuum field theory based approaches\cite{PhysRevB.27.5592,PhysRevB.47.11093}. Early perturbative methods utilized either the impurity concentration, or the deviation from a mean mass as a small parameter. Later approaches were  based on the coherent potential approximation (CPA) and the average T-matrix approximation (ATA). More recently, Ghosh et al. \cite{PhysRevB.66.214206} developed the itinerant coherent-potential approximation (ICPA) which satisfies translational invariance, unitarity, and analyticity of physical properties. 
The ICPA has two additional advantages. First, it can capture the physics of multi-site correlations. Second, it can incorporate both mass and spring disorder simultaneously. In this connection, the ICPA is one of the most successful extensions of the CPA to predict the vibrational density of states of realistic binary alloy systems. Nevertheless, the ICPA is not able to capture the AL of phonons.

Approximate theories such as the CPA or the ATA may be used to get a qualitative insight. However, these are often based on uncontrolled approximations, and their region of validity is always in question. This is where numerically exact methods such as exact diagonalization (ED)\cite{RevModPhys.44.127,PhysRev.154.802} and transfer matrix method (TMM)\cite{0295-5075-97-1-16007} prove their mettle and provide very useful benchmarks for approximate theories. Recently, Monthus and Garel \cite{PhysRevB.81.224208} use ED for relatively large system sizes to investigate the localization of phonons in mass-disordered systems. Using finite size-scaling methods for the low-frequency part of the spectrum, they show that the single-parameter scaling theory of localization, originally developed for electronic systems applies to phononic systems as well. Pinski et al. \cite{0295-5075-97-1-16007} employ the TMM to obtain the mobility edge as a function of mass and spring disorder in three-dimensional systems. They find a close correspondence between the electronic and phonon systems. The main drawback of ED and TMM is that their computational expense scales exponentially with system size.

Despite extensive investigations over decades, a method that fulfills all of the following set of requirements has not yet been developed: (1) The method should systematically approach the thermodynamic limit. (2) It should reproduce exact diagonalization results for both the main vibrational spectrum and the impurity modes.  (3) It should be applicable over the full alloy regime, i.e., for all defect concentrations. (4) It should be able to handle both mass (diagonal) and spring (off-diagonal) disorder on an equal footing. (5) It should capture the AL of phonons, including the dependence of the mobility edge on the disorder. (6) It should be relatively computationally inexpensive in order to be useful for investigations of phonon localization in real materials, which necessarily involve multiple branches, and mass as well as spring disorder.

The lack of a single method satisfying all the criteria mentioned above for phononic systems motivates us to adapt the dynamical cluster approximation (DCA) 
and the typical medium DCA (TMDCA) for disordered phononic systems to capture the Anderson localization of phonons since these methods have been shown to work extraordinarily well in electronic systems \cite{RevModPhys.77.1027,PhysRevB.89.081107,PhysRevB.90.094208,PhysRevB.94.224208} .

The main difficulty inhibiting the development of such a method for the study of 
Anderson localization of phonons lies in finding a single particle order parameter to characterize the Anderson transition in disordered phononic systems. Recently, a typical medium theory (TMT) \cite{epl_order_parameter} for electronic systems proposes the local density of states (LDOS) as an appropriate quantity to look at for the study of Anderson localization of electrons. The local density of states, defined as $\rho_l(\omega)=\sum_n \delta(\omega-\omega_{n}) |\psi_n(l)|^2$,  changes from continuous to discrete upon the system transiting from an itinerant to a localized state. On the insulating side of the transition, the spectrum consists of delta functions. Here, the typical value of the LDOS vanishes, whereas the globally averaged density of states (ADOS) does not, nor is it critical at the Anderson transition. Hence, the TMT adopts the typically averaged DOS (TDOS), as an order parameter for the study of the Anderson localization of electrons.  In spite of the success of the TMT in describing localized electron states, it suffers shortcomings due to its single-site character. For example, the TMT does not provide a proper description of the critical behavior of the Anderson localization transition in three dimensions for disordered electronic systems since it is not able to capture the effects of non-local coherent back-scattering.

Recently, an extension of the TMT that includes non-local dynamical correlations, called the typical medium dynamical cluster approximation(TMDCA),\cite{PhysRevB.89.081107} has been developed for disordered electronic systems. It incorporates the typical medium within the dynamical cluster approximation (DCA) scheme. The TMDCA possesses all features of a successful cluster theory such as the systematic incorporation of non-local correlations, and it captures the critical behavior of the Anderson localization transition including the correct value of critical disorder strength and re-entrant behavior of the mobility edge. 
  
The present TMDCA method for electronic systems utilizes the fact that the LDOS is a continuum in the metallic state; whereas it is composed of a set of delta-functions in an insulator, so that the typical value of the LDOS, averaged over disorder locations, is zero. This same idea is equally applicable to phonons or the localization of any propagating waves. So, we can consider that the typical value of the LDOS remains a valid order parameter for phononic systems. Based on this concept, we establish a TMDCA formalism for the study of Anderson localization of phonons.  

We end this introduction with two questions: (1) (a) How well do the DCA and the TMDCA formalisms do when compared with exact methods like ED and TMM? 
(b) To what extent are the requirements of a successful method, that are mentioned above, fulfilled by the DCA and the TMDCA? (2) What new insights into the localization of phonons does the calculation of typical density of states give? These questions will be addressed at the appropriate places in the manuscript. In the present work, we have focused on diagonal mass-disorder. Thus, the issues of spring disorder and anharmonicity have not been considered and are reserved for future studies.  In the following section, we describe a model for a mass disordered lattice within the harmonic approximation and the formalism employed to solve the model.

\section{METHOD}
The Hamiltonian for the ionic degrees of freedom of a  disordered lattice in the harmonic approximation can be written in terms of momentum ($p$) and displacement ($u$) operators, as
\begin{equation}
H=\sum_{\alpha i l}\frac{{p^2_{i\alpha}(l)}}{2M_{i}(l)} + \frac {1}{2} \sum_{\alpha\beta ll^\prime ij} \Phi^{\alpha\beta}_{ij}(l,l^\prime)u_{\alpha}^i(l)u_{\beta}^j(l^\prime)\,,
\end{equation}
where $p_{i\alpha}(l)$ and $u^i_{\alpha}(l)$ represents, respectively,  the momentum and the displacement (from the equilibrium position) of a site $i$ belonging to the unit cell $l$ along  the Cartesian coordinate $\alpha=(x,y,z)$ direction. The index $i$ runs from $1$ to $N_{\rm cell}$ where the latter denotes the 
number of atoms in the basis.  We assume that the force-constant tensor,
$\Phi$, is a function of $\left| {\mathbf{R}_i}(l)- {\mathbf{R}_j}(l^\prime)\right|$, where $\mathbf{R}_i(l)$ is the position of ion $i$ in unit cell $l$.

The retarded displacement-displacement Green's functions, 
\begin{equation}
iD^{ij}_{\alpha\beta}(l,l^\prime,t)=\langle\langle\, u^{i}_{\alpha}(l,t);u^{j}_{\beta}(l^\prime,0)\, \rangle\rangle
\label{eq:gf}
\end{equation}
corresponding to the above Hamiltonian can be obtained using their 
frequency dependent counterparts given by (see Appendix A for details) the solution of the following coupled linear equations:
\begin{align}
M_i(l) \omega^2 D_{\alpha\beta}^{ij}(l,l^\prime,\omega) &= \delta_{\alpha\beta}\delta_{ll^\prime}\delta_{ij}\nnu \\
&+\sum_{\gamma,l^{\prime\prime}j^\prime} {\Phi^{\alpha \gamma}_{i j^\prime}(l, l^{\prime\prime})} D_{\gamma \beta}^{j^\prime j}(l^{\prime\prime}, l^\prime,\omega)\,. 
\label{eq:eom}
\end{align}
With a single composite index, $\lambda=(\alpha,l,i)$, we can write the above equation in a matrix representation and obtain a formal solution for the Green's function (Eq.~\ref{eq:gf}) as 
\begin{equation}
M_0{\hat{D}}(\omega)=\left[\omega^2\mathbb{1} - \hat{\Phi}M^{-1}_0 - \omega^2 {\hat{V}}\right]^{-1}\,,
\label{eq:gfsol}
\end{equation}
where $M_0$ includes the masses of the ions in the unit cell of the clean lattice with respect to which the mass
'disorder potential', ${\hat{V}}$, is given as
\begin{equation}
\left({\hat{V}}\right)_{\lambda,\lambda^\prime} = \left(1 - M_\lambda M^{-1}_{0 \lambda'}
\right)\delta_{\lambda,\lambda^\prime}\,.
\label{eq:dispot}
\end{equation}
Note that the masses have been assigned a Cartesian index purely for notational convenience, {\it i.e.} the mass of the $i^{\rm th}$ atom in the $l^{\rm th}$ unit cell does not, naturally, depend on $\alpha$, the direction.

In this work, we consider an isotropic simple cubic lattice with a monoatomic basis 
($M_{0 \lambda}= M_0$) and a spring constant tensor $\Phi$ truncated at nearest-neighbors:
\begin{equation}
\Phi^{\alpha\beta}(l,l^\prime) = \delta_{\alpha\beta}(\Phi_D\delta_{l,l^\prime}
 + \Phi_{nn}\delta_{\bR_{l^\prime},\bR_l+\vec{\delta}})\,,
 \label{eq:phiform}
\end{equation}
where $\Phi_D$ and $\Phi_{nn}$ are the diagonal, and the nearest neighbor component of the tensor, respectively, and $\vec{\delta}$ is a vector connecting a site to its nearest neighbors.

We consider two kinds of mass disorder in Eq.~\ref{eq:dispot}, namely (1) binary isotopic disorder, where the random masses $M_\lambda$ are either $M_{\rm imp}$ or $M_0$ with  concentrations $c$ and $(1-c)$ , respectively, and (2) a uniform (box) disorder, where $(1-M_\lambda/M_0)\in [-V,V]$ with equal probability for any value in that interval and $0<V\leq 1$ representing the strength of disorder . Binary isotopic disorder is a special case of binary disorder, since the latter may involve substitutions that may induce spring disorder in addition to mass disorder. Most experimental studies involve disorder in a binary alloy. Hence, we perform calculations for this disorder distribution. However, a comprehensive validation of the numerical schemes requires us to compare our results with the available results for the box distribution.  Thus, the two distributions are needed to complete our study. 

In the absence of mass disorder, i.e ${\hat{V}}=0$, corresponding to a clean, monoatomic lattice, all ionic masses are identical, hence $M_i(l)=M_0$ and $i=1$ for all $l$ lattice sites.   In such a case, the system is translationally invariant, hence transforming to $\bk$-space using 
\begin{displaymath}
 M_0D^{(0)}_{\alpha\beta}(l,l^\prime,\omega) = \sum_\bk D^{(0)}_{\alpha\beta}(\bk,\omega)
 e^{i\bk\cdot(\bR_l - \bR_{l^\prime})}\,,
\end{displaymath} 
Eq. \ref{eq:eom} simplifies to 
\begin{equation}
{\bar{D}}^{(0)}(\bk,\omega) = \left[\omega^2\mathbb{1} - \bar{F}(\bk)\right]^{-1}
\end{equation}
where the `bar' represents a matrix in the Cartesian basis
(e.g. $3\times 3$ in three dimensions), and
$\bar{F}(\bk)$ is related to ${\hat{\Phi}}$ through
\begin{displaymath}
 (\bar{F}(\bk))_{\alpha\beta}=\sum_{l^\prime} \frac{\Phi_{\alpha\beta}(l,l^\prime)}{M_0} e^{i\bk\cdot(\bR_l - \bR_{l^\prime})}\,.
\end{displaymath} 
Thus, with the specific form for $\Phi$ given by Eq.~\ref{eq:phiform},
the Green's function in the clean limit reduces to
\begin{equation}
{\bar{D}}^{(0)}(\bk,\omega) = \left(\omega^2 - \omega_\bk^2 \right)^{-1}\mathbb{1},
\label{eq:clean}
\end{equation}
where the dispersion is given by
\begin{equation}
\omega_\bk^2 = \omega_0^2\left(\sin^2 \frac{k_x}{2}+\sin^2 \frac{k_y}
{2}+ \sin^2 \frac{k_z}{2}\right),
\label{eq:disp}
\end{equation}
with $\omega_0=\sqrt{4\gamma/M_0}=1$ being our unit of energy and $\gamma=-\Phi_D = 6\Phi_{nn}$; the latter equality stems from sum rules
that need to be satisfied by the spring constant tensor.
The choice of $\omega_0=1$ implies that the bandwidth of the non-interacting spectrum is $\sqrt{3}$. 
Since all the branches have identical dispersion,  we will drop the branch index ($\alpha$) henceforth
in this work.
Thus, Eq.~\eqref{eq:gfsol} may be written as a Dyson equation:
\begin{equation}
{\hat{D}}^{-1}(\omega) = \left({\hat{D}}^{(0)}(\omega)\right)^{-1} - \omega^2{\hat{V}}\,.
\label{eq:dyson}
\end{equation}

The connection to disordered electronic systems can now be made.  The non-interacting electronic Green's function in a clean lattice is given by $G^{(0)}(\bk,\omega)=(\omega^+ - \epsilon_\bk)^{-1}$, where $\epsilon_\bk=-2t(\cos(k_x) +\cos(k_y)+\cos(k_z))$ is the electronic dispersion in a cubic lattice with nearest-neighbor hopping $t$. By noting that the phonon dispersion (Eq.~\eqref{eq:disp}) can be mapped to the electronic dispersion through $\omega_\bk^2= 6\gamma/M_0+\gamma\epsilon_\bk/(M_0 t)$, the similarity
between $G^{(0)}$ and $D^{(0)}$ (Eq.~\eqref{eq:clean}) becomes immediately clear. 

A major difference between the localization of phonons and electrons emerges from the form of the Dyson equation. In the electronic case, ${\hat{V}}$ represents site-disorder and the Dyson equation reads ${\hat{G}}^{-1}(\omega) = \left({\hat{G}}^{(0)}(\omega)\right)^{-1} - {\hat{V}}$, while in the phonon case, the perturbation term is $\omega^2 {\hat{V}}$ (Eq.~\eqref{eq:dyson}), which creates a significant difference in
the localization of phonons {\it vs} electrons. For example,
localizing low energy acoustic modes should be almost impossible
because the modulating factor of $\omega^2$ implies that the
disorder potential becomes vanishingly small at low energies, hence leaving the acoustic modes almost unperturbed. The implication for high-frequency modes is also clear: the disorder potential increases without bound; hence high-frequency modes are expected to get localized even
for relatively weak disorder. Further differences will be pointed out in the results section.
  
There are several methods to solve the Dyson equation (Eq.~\eqref{eq:dyson}).  Diagrammatic methods employing an infinite resummation of a certain class of diagrams are one choice\cite{PhysRev.181.1006}. The CPA, which reduces the lattice problem to an effective single-site problem, is another. Alternatively, one can choose a finite system with periodic boundary conditions and solve for the Green's function exactly. Each of these methods has specific advantages and disadvantages.  For example, the diagrammatic methods are often uncontrolled approximations and may violate sum rules and/or yield unphysical spectra. 

Finite system calculations, though exact, suffer from a large computational expense. Hence, a method is needed that is computationally feasible, systematically approaches the thermodynamic limit and is fully causal. The dynamical cluster approximation (DCA) is one such method. It has been applied very successfully to investigate a variety of fermionic and bosonic models. In this work, we extend the DCA to study phonons in mass-disordered systems. We now describe the DCA for phonons in some detail below.
 
\subsection{Dynamical cluster approximation (DCA) for phonons}
Jarrell et al. \cite{PhysRevB.63.125102} introduced the DCA as an extension of the dynamical mean field approximation (DMFA) through the inclusion of non-local spatial correlations. DCA systematically incorporates non-local correlations by mapping the original lattice problem onto a periodic cluster of size $N_c\sim L_c^d$ where $L_c$ is the linear size of the cluster and $d$ is the dimension of the lattice. The periodic cluster is embedded into a self-consistent effective medium which is characterized by a non-local hybridization function $\Gamma(\mathbf K,\omega)$.  The effective medium is constructed via algebraic averaging over disorder configurations.  Hence, spatial correlations up to a range $\xi \simeq L_c$ are taken into account accurately, while the longer length scale physics is treated at the mean-field level.  In this formulation, it is assumed that the momentum dependence of the hybridization function is weak.   An algorithm that implements the DCA for solving Eq.~\eqref{eq:dyson} for phonons is given below:

\noindent
1. The computational scheme  begins with an initial guess  for the hybridization function $\Gamma_{\rm old}(\mathbf K,\omega)$. Such a guess can be obtained either through a previous calculation or through a coarse-graining of the non-disordered Green's function (Eq.~\eqref{eq:clean}):
\begin{equation}
\Gamma_{\rm old}(\mathbf K,\omega) = \omega^2 - 
\bar \omega_\mathbf K^2 - \left(\sum_\bkt D^{(0)}(\bK+\bkt,
\omega)\right)^{-1},
\end{equation}
where $\bkt$ runs over the momenta of the cell centered at the cluster momentum 
$\bK$, and $\bar \omega_\bK^2$ is the coarse-grained dispersion given by
\begin{equation}
\bar \omega_\bK^2= \frac{N_c}{N}\sum_{\bkt} \omega_{\bK+\bkt}^2\,,
\end{equation}
where $\omega_{\bk}^2$  is given in Eq.~\eqref{eq:disp}.

\noindent
2. The hybridization function is used to calculate the cluster excluded Green's function ${\mathcal D}(\bK,\omega)$ as
\begin{equation}
{\mathcal D}(\bK,\omega)=\frac {1}{\omega^2-\bar \omega_\bK^2-\Gamma_{\rm old}(\bK,\omega)}\,.
\end{equation}

\noindent
3. 
The cluster excluded Green's function in momentum space
 is Fourier transformed to real space: 
\begin{align}
M_0{\mathcal D}(l,l^\prime,\omega)  =\sum_{\bK} {\mathcal D}(\bK,
\omega)
 \exp\left(i \bK\cdot(\bR_l-\bR_{l^\prime})\right).
\end{align}

\noindent  
4. Next, we generate a large number of configurations of the disorder potential (${\hat{V}}$) for a given distribution, namely binary isotopic or box disorder.

\noindent
5. For each disorder configuration $\hat V$, the mass-weighted Dyson equation is used to compute the cluster Green's function, given by
\begin{align}
& D^c(l,l^\prime,\omega) = \nnu \\
& \sqrt{1-(\hat V)_l} {\left[ ({\hat{\mathcal D}}(\omega))^{-1}- \omega^2{\hat V}  \right]}_{ll^\prime}^{-1}
\sqrt{1-(\hat V)_{l^\prime}}\,,
\label{eq:mwdyson}
\end{align}
 which is then averaged over all disorder configurations:
\begin{equation}
D^c_{\scriptscriptstyle{\rm DCA}}(l,l^\prime,\omega)  =  \bigg\langle  D^c(l,l^\prime,\omega)  \bigg\rangle
\label{eq:dcaave}
\end{equation}
where $\langle ... \rangle$ denotes an algebraic average. As explained in Appendix B, the mass-weighting is essential in order to ensure a proper normalization of the spectral functions in the presence of disorder. 
In practice, we have generated about 600-1000 disorder configurations for each simulation, and have verified the robustness of our results with respect to the number of configurations.

\noindent
6. The average cluster Green's function obtained in Eq.~\eqref{eq:dcaave} is Fourier transformed to momentum space, and then used to compute the 
coarse-grained lattice Green function: 
\begin{align}
& D^{\scriptscriptstyle CG}(\mathbf K,\omega)  = \nnu \\
& \frac{N_c}{N}\sum_{\bkt} \left[  \left(D^{c}_{\scriptscriptstyle{\rm DCA}}(\bK,\omega)\right)^{-1}  + 
\Gamma_{\rm old}(\bK,\omega)-\omega_{\bK+\bkt}^2 + \bar{\omega}_\bK^2 \right]^{-1}\,.
\label{eq:cg}
\end{align}

The disorder averaged spectral function, termed the ADOS may be defined as
\begin{equation}
{\rm ADOS}(\omega^2)=-\frac{2\omega}{N_c\pi}{\rm Im}
\sum_\bK D^c_{\scriptscriptstyle{\rm DCA}}(\bK,\omega)\,.
\label{eq:dca-dos}
\end{equation}
\noindent
7. A new hybridization function is found through 
\begin{align}
\Gamma_{\rm new}(\bK,\omega)&=  \Gamma_{\rm old}(\bK,\omega)+
\nnu \\ 
& \xi \left[\left(D^{\scriptscriptstyle CG}(\bK,\omega)\right)^{-1} -\left(D^{c}_{\scriptscriptstyle{\rm DCA}}(\bK,\omega)\right)^{-1}\right]
\label{eq:sc}
\end{align}
where $\xi$ is a linear mixing parameter used for improving the convergence.

Self-consistency is achieved when $|| \Gamma_{\rm new} (\bK, \omega)-\Gamma_{\rm old}(\bK, \omega)||$ 
reaches numerical
tolerance (in practice, about 0.005).  We have checked that such
a condition is sufficient to obtain converged Green's
 functions and self-energies.
If the self-consistency condition is satisfied, the iterations end, else we impose $\Gamma_{\rm old}=\Gamma_{\rm new}$ and go back to step-2. In practice, we add a small imaginary broadening factor ($\omega\rightarrow \omega+\iota\eta; \eta\sim 10^{-3}$) to real frequencies for accelerating convergence.

Though the DCA possesses several advantages over the CPA, both are unable to capture the Anderson localization. The arithmetic averaging used for computing the cluster Green's function  (Eq.~\eqref{eq:dcaave}) in step-5 leads to this inability. The typical medium DCA developed for electronic systems has been demonstrated to capture Anderson localization. We describe the extension of DCA to TMDCA for phonons below.

\subsection{Typical Medium Dynamical Cluster Approximation (TMDCA) for phonons}
As mentioned above, the DCA employs algebraic averaging over disorder configurations, while in the TMDCA, the effective medium is constructed via geometric averaging. The ansatz for computing the typical density of
   states remains the same as in the electronic case, namely:
\begin{align}
\rho^{c}_{\rm typ}(\bK,\omega) =\exp \left(\frac{1} {N_{c}} \sum_{l=1}^{N_{c}}\left\langle \ln \rho^{c} (l,\omega)
\right\rangle \right) \nnu \\
\times \left\langle \frac {\rho^{c}(\bK,\omega)} {\frac{1}{Nc}\sum_{l}\rho^{c}(l,\omega)}\right\rangle
\label{eq:typdos}
\end{align}
where
\begin{align}
\rho^c(l,\omega) &=-\frac {2 \omega}{\pi}{\rm Im} \,D^c(l,l,\omega) \nnu \\
\rho^{c}(\bK,\omega) & = -\frac {2 \omega} {\pi}  {\rm Im}\, D^c(\bK,\omega) \nnu
\end{align}
are the local and momentum dependent spectral functions respectively, computed from the unaveraged cluster Green function $D^{c} (l,l^\prime,\omega)$ (Eq.~\eqref{eq:mwdyson}). 

The disorder-averaged typical Green's function can be calculated from the typical density of states (Eq.~\eqref{eq:typdos}), using the Hilbert transform as  
\begin{equation}
D_{\rm typ}^{c}(\mathbf K,\omega) = {\mathcal P} \int d\omega^\prime \frac {{\rho^{c}_{\rm typ} (\bK,\omega^\prime)}} {\omega^2-\omega^{\prime 2}}
-i \frac {\pi}{2\omega} {{\rho^{c}_{\rm typ} (\omega}})\,,
\label{eq:ht}
\end{equation}
and the corresponding typical density of states, termed the TDOS
is given by
\begin{equation}
{\rm TDOS}(\omega^2)=-\frac{2\omega}{N_c\pi}{\rm Im}
\sum_\bK D^c_{\scriptscriptstyle{\rm typ}}(\bK,\omega)\,.
\label{eq:def-tdos}
\end{equation}

\begin{figure}[h!]
\centerline{\includegraphics[scale=0.36]{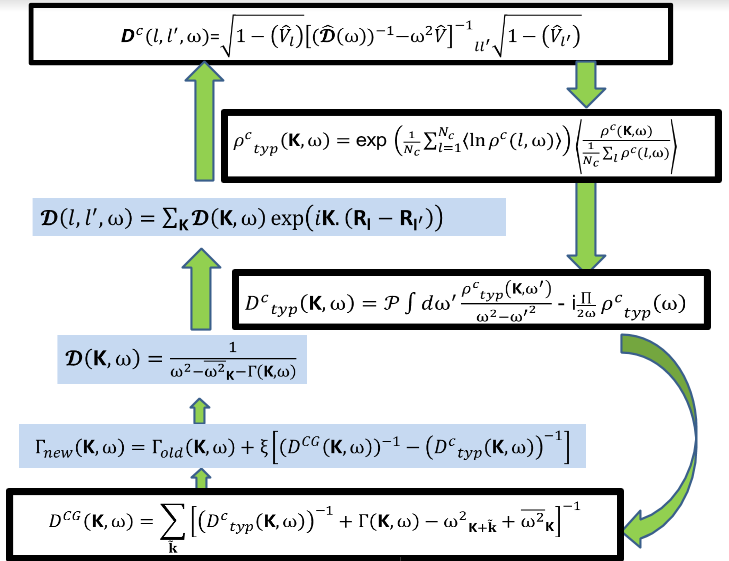}}
\caption {Self-consistency loop of the TMDCA for phonons.}
\label{fig:flowchart}
\end{figure}

The TMDCA implementation is almost identical to that of the DCA, except that the typical Green's function is obtained by combining Eqs.~\eqref{eq:mwdyson}, ~\eqref{eq:typdos} and ~\eqref{eq:ht} and in
Eqs.~\eqref{eq:cg} and \eqref{eq:sc}, the $D^c_{\scriptscriptstyle{\rm DCA}}$ is
 replaced by $D^c_{\rm typ}$. 
The flowchart of the algorithm is presented in Fig.~\ref{fig:flowchart}.
Apart from the typical Green's function, $D_{\rm typ}^{c}$, an average Green's function, denoted by $D_{\rm typ}^{\rm ave} $ can also be computed within the TMDCA using Eq.~\eqref{eq:dcaave} in the final iteration of the TMDCA self-consistency cycle. An interpretation of such a Green's function is that it yields the physical density of states, while the typical density of states acts as an order parameter for the
Anderson localization transition.

The rest of the  paper is organized as follows: We will validate the DCA and TMDCA against exact diagonalization and transfer matrix method respectively in section~\ref{bench}. Subsequently in section~\ref{sec:tmdca}, the typical density of states, computed through TMDCA,  is used to discuss the physics of phonon localization. Conclusions are presented in the final section.

\section{Benchmarking DCA and TMDCA}
\label{bench}
The first step to establish any new method is to
benchmark it against previous exact results. This will be  the objective of this section. The DCA and TMDCA benchmarks are established separately in subsections~\ref{bench_dca} and ~\ref{bench_tmdca} respectively.

\begin{figure}[t]
\includegraphics[scale=0.4]{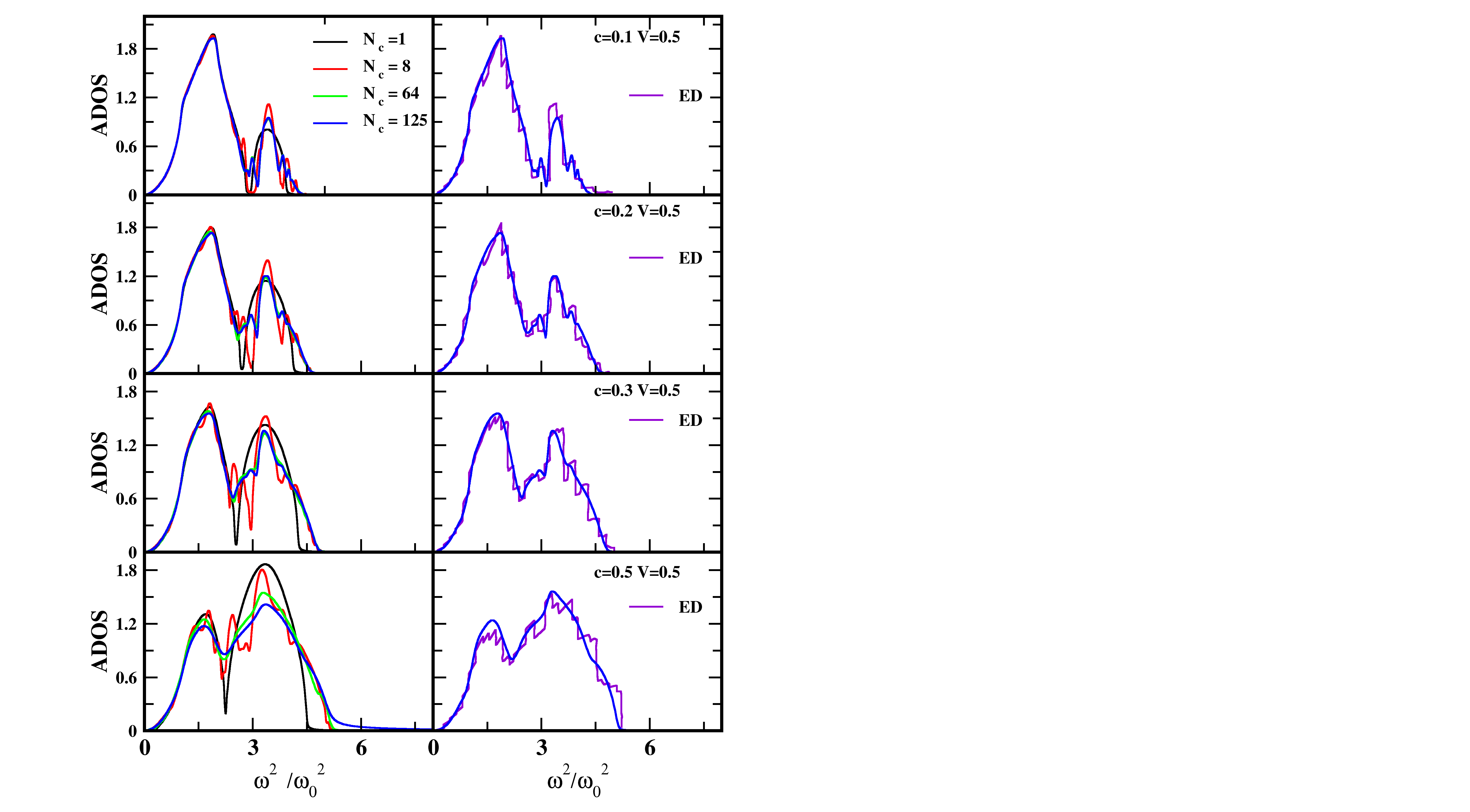}
\caption{(Color online) Comparison of the density of states obtained from the DCA and exact diagonalization (ED) methods for a binary isotopic alloy system in three dimensions at
a fixed mass ratio ($V=0.5$) and various values of concentration ($c$). The left panel shows the DCA results for increasing impurity concentration $c$ (from top to bottom). Each panel illustrates the evolution of the spectrum with increasing cluster size. The right panel shows a comparison of the $N_c=125$ DCA result with ED (data from Ref. \onlinecite{PhysRev.154.802} ) for the same parameters. The agreement between the DCA and ED is seen to be excellent, whereas there is strong disagreement between $N_c=1$ results and ED results.}
\label{fig:compare_ED_3D}
\end{figure}
\subsection{Dynamical Cluster Approximation}
\label{bench_dca}


Fig.~\ref{fig:compare_ED_3D} shows
a direct comparison of the density of states obtained from the DCA with results from exact diagonalization (ED) \cite{PhysRev.154.802} for a binary isotopic alloy system in three dimensions at various values of disorder potential ($V$) and concentrations ($c$). 
The disorder averaged density of states can be obtained from the DCA cluster Green's function $D^c_{\scriptscriptstyle
{\rm DCA}}$ (Eq.~\eqref{eq:dcaave}) and
is given by Eq.~\eqref{eq:dca-dos}.

The DCA calculations have been performed
for a simple-cubic lattice with different cluster sizes, namely $N_c=1,8,64,125$.
In the ED calculations \cite{PhysRev.154.802}, the DOS was calculated for a $6\times 6 \times 25$ randomly disordered simple-cubic lattice. The left panels of Fig.~\ref{fig:compare_ED_3D} show the evolution of the spectrum
with increasing (from top to bottom) concentration ($c$) of light impurities (with $M_{imp}=M_{host}/2$,
hence $V=0.5$).  The two-peaked structure of the spectrum, seen for all concentrations, is reflective of the binary mass distribution. The spectral weight of the higher frequency band is seen to grow with increasing $c$, while the low-frequency band shrinks. For $c\gtrsim 0.5$, the system may be viewed as the dual
of the original system, i.e., a binary alloy with a lighter host and heavier impurities. 
The transfer of spectral weight is natural since lighter impurities should have higher characteristic frequencies.

The DCA for a single-site cluster ($N_c=1$) reduces to the CPA. The left panels of
Fig.~\ref{fig:compare_ED_3D}
also shows that results from the CPA are quite different from those at higher $N_c$, thus emphasizing the need to incorporate non-local dynamical correlations.  Nevertheless, we note that the CPA roughly captures the overall shape. There are two problems, however. At the lowest frequencies, the CPA spectral function exhibits a gap, while the DCA spectra (for higher $N_c=64,125$) do not. In fact, even the $N_c=8$ spectrum is gapped, albeit with a smaller gap as compared to the CPA.  The reason for this spurious gapped behavior is that the correct sum rules are obeyed only in the thermodynamic limit. The second problem is that in the high-frequency region, the CPA spectrum comprises an almost separated impurity band with a cusp-like non-analytic feature. This feature is again in contrast with results of higher $N_c$,  which shows that the spectrum is continuous and broad.  Moreover, we observe that results for $N_c=64$ and $N_c=125$ are hardly different for all concentrations,
 suggesting that the convergence with respect to increasing in cluster size is achieved for a cluster as small as $4 \times 4\times4$.

The right panels of Fig.~\ref{fig:compare_ED_3D} show a direct comparison of results using the DCA at the highest $N_c=125$ of the corresponding left panel with ED results~\cite{PhysRev.154.802}. In general, the computational expense in ED depends on many factors; like the number of frequencies, the length of the lattice and also on the number of atoms in a cross section of the lattice. We consider ED results from Ref \onlinecite{PhysRev.154.802}, where they use a $6\times 6 \times 25$ lattice  and a Strum sequence method.  Clearly, the agreement between the ED and DCA, even considering the fine structure of the ED results, is rather good. Thus, the DCA is not only far less expensive than the ED but is also able to yield a smooth and continuous spectrum. Furthermore, the DCA converges to the exact, thermodynamic limit result far more rapidly than the ED, which achieves convergence for much larger system sizes ($6\times 6 \times 25$). Thus, our DCA scheme can efficiently calculate the average vibrational spectra in three dimensions for arbitrary values of impurity concentrations and disorder potential.   

Since the DCA is non-perturbative, it is applicable over the entire alloy regime, $c\in [0,1]$, which has been a significant limitation of perturbative theories of alloys \cite{PhysRevB.10.1190,PhysRev.131.163,taylor}.
A cluster approach developed by Myles and Dow\cite{PhysRevB.19.4939} also incorporates non-local correlations. However, this method is limited by a restriction on the combined choice of concentration ($c$) and cluster size ($N_c$), which have to obey the relation, $cN_c$=integer, akin to supercell-based calculations. The DCA does not suffer from this restriction, which makes it possible to access any impurity concentration for a given cluster size. Another drawback of Myles' cluster method as compared to the DCA is that the effective medium is described within the CPA. As a result, the bandwidth of the local impurity mode obtained from Myles' calculations is too narrow.  

The excellent benchmark obtained thus far implies that the DCA scheme for phonons with increasing cluster size ($N_c$) can efficiently predict vibrational spectra for
disordered systems. 
Nevertheless, the DCA  is not able to capture Anderson localization\cite{PhysRevB.89.081107,PhysRevB.63.125102} of phonons. In order to incorporate the physics of localization, we utilize the TMDCA method,
described in section II. The following sub-section describes the validation of the TMDCA through a direct comparison with the transfer matrix method.

\begin{figure}[b]
\includegraphics[width=\linewidth,trim=1cm 0.8cm 2cm 3.7cm,clip]{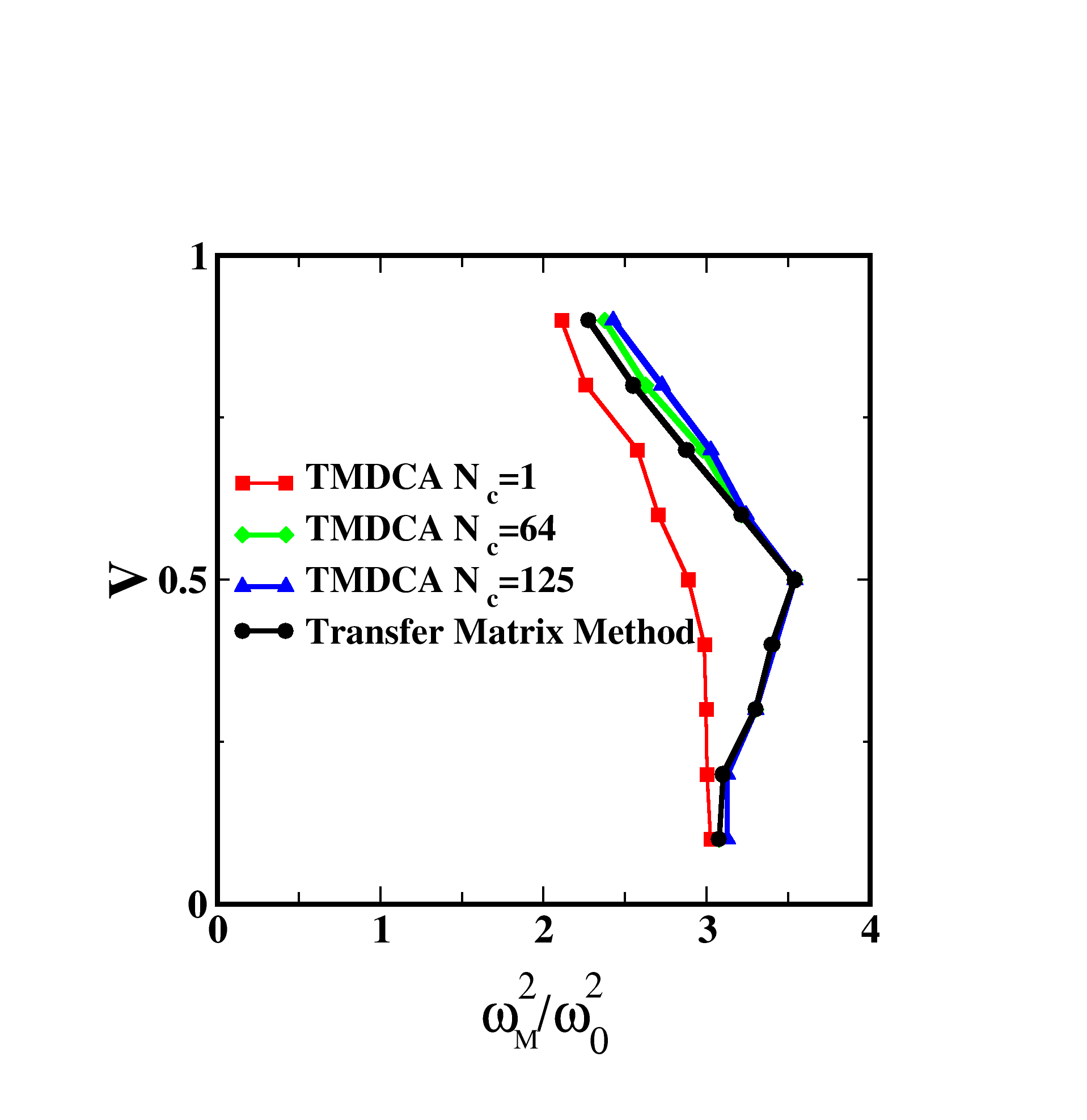}
\caption{A comparison of the mobility edges ($\omega_M$) in the phonon spectrum in three dimensions for a box-distribution, obtained from the transfer matrix method \cite{0295-5075-97-1-16007} against TMDCA. Results from the latter 
for the larger clusters agree excellently with the TMM results.}
\label{fig:mobility_edge_phase}
\end{figure}

\subsection{Typical Medium Dynamical Cluster Approximation}
\label{bench_tmdca}

A striking feature of disordered systems in three dimensions is the existence, in the density of states, of a mobility edge \cite{0022-3719-20-21-008}, which is defined as the energy separating localized and itinerant states.  Experimental measurements
of the mobility edge are feasible as demonstrated for ultracold atoms in a disordered potential created by laser speckles~\cite{mobilityedge}.  Within the TMDCA, the mobility edge is determined using the band-edges of the typical density of states (TDOS), since the latter is non-zero only for extended states. For a box disorder distribution, defined as $P_{V}(V_l) = \Theta(V-|V_l|)/2V$ where $V_l=(1-M_l/M_0)$ and $0<V\leq 1$, where $l$ is the site index, and $V$ is the width of the distribution that represents the strength of disorder, the 
mobility edge determined using TMDCA is compared against exact transfer matrix method results in Fig.~\ref{fig:mobility_edge_phase}. 
The agreement between results from the TMM (black circles) and the TMDCA for $N_c=64$ (green diamonds) and $125$ (blue triangles) is excellent. Such a result is not surprising, since the TMDCA, for three-dimensional {\em electronic} disordered systems,  agrees very well with the kernel polynomial method and the transfer matrix method\cite{PhysRevB.89.081107}. 

For $V\gtrsim 0.5$, the TMM results and likewise those from TMDCA exhibit a re-entrant transition with increasing disorder, in parallel with the behavior in disordered electronic systems \cite{PhysRevB.89.081107}.
However, an important difference is that beyond a critical disorder, all the states in the electronic system become localized; while in the phonon case, a finite fraction of the low-frequency states remain extended.  
In analogy with the electronic case, the re-entrance transition seen in the TMDCA results in Fig.~\ref{fig:mobility_edge_phase} has the following explanation: Very low disorder induces states outside the band-edge that merge with the continuum through hybridization. At intermediate levels of disorder, isolated  localized modes (analogous to deep trap states) appear beyond the band-edge, which nevertheless hybridize with each other and the extended states on the band-edge, and thus transform into extended states. We note that such a hybridization requires inter-site correlations, that are missing from a single site theory ($N_c=1$) such as the TMT, and hence a blue shift of the mobility edge (seen in the TMDCA results of Fig.~\ref{fig:mobility_edge_phase}) is not captured by the single-site theory.  However, with increasing disorder, states at the band edges begin to get localized, and hence the mobility edge undergoes a re-entrance crossover. 

The failure of single-site theories ($N_c=1$, red squares in Fig.~\ref{fig:mobility_edge_phase}), as evidenced by the significant disagreement with TMM results involves two factors: (i) The TMM mobility edge initially blue shifts with increasing disorder, while the $N_c=1$ result red shifts monotonically.  (ii) The TMM as well as the TMDCA results for higher disorder strengths ($V\gtrsim 0.5$), clearly show a re-entrant transition, which the single-site theory completely misses. Likely, this is due to the fact that the $N_c=1$ calculation is a single-site theory and hence does not incorporate non-local coherent back-scattering effects; 
although it does include strong localization effects induced by deep trapped states. 



  
With these results for the DCA and TMDCA, the question 1(a) posed at the end of the introduction is fully answered. Both DCA and TMDCA do yield excellent agreement when compared to exact methods. Now, we move to a discussion of  results on the Anderson localization of phonons.

\section{Results from TMDCA} 
\label{sec:tmdca}

In disordered electronic systems the typical density of states, given by Eq.~\eqref{eq:def-tdos}, may be used as an order parameter for the Anderson localization transition~\cite{epl_order_parameter,PhysRevB.89.081107}.
While the physical observable is still the {\em arithmetically} averaged density of states (${\rm ADOS}(\omega)$), the ${\rm TDOS}(\omega)$ yields a mobility edge that separates localized and extended states.  Within the TMDCA, the ${\rm ADOS}(\omega)$ is computed from $D^{\rm ave}_{\rm typ}(\bK,\omega)$, which, as explained
in Section II, carries information about the 
typical medium within which the cluster is embedded.
The hybridization function connecting the cluster with
the host is known \cite{RevModPhys.77.1027} to decay as a function of increasing cluster size
 as $\sim {1/N_c^2}$. Hence, the ${\rm ADOS}(\omega)$
 computed within TMDCA must coincide with the corresponding quantity computed within the DCA in the thermodynamic limit. In practice, we find that even at $N_c=64$, the two are almost identical. This is shown in Fig.~\ref{fig:fig6} of Appendix C. In what follows, we will discuss results
 for the average and the typical density of states, computed through TMDCA for box and binary disorder distributions. 

\subsection {Box disorder}
\begin{figure}[t]
\centerline{\includegraphics[scale=0.6]{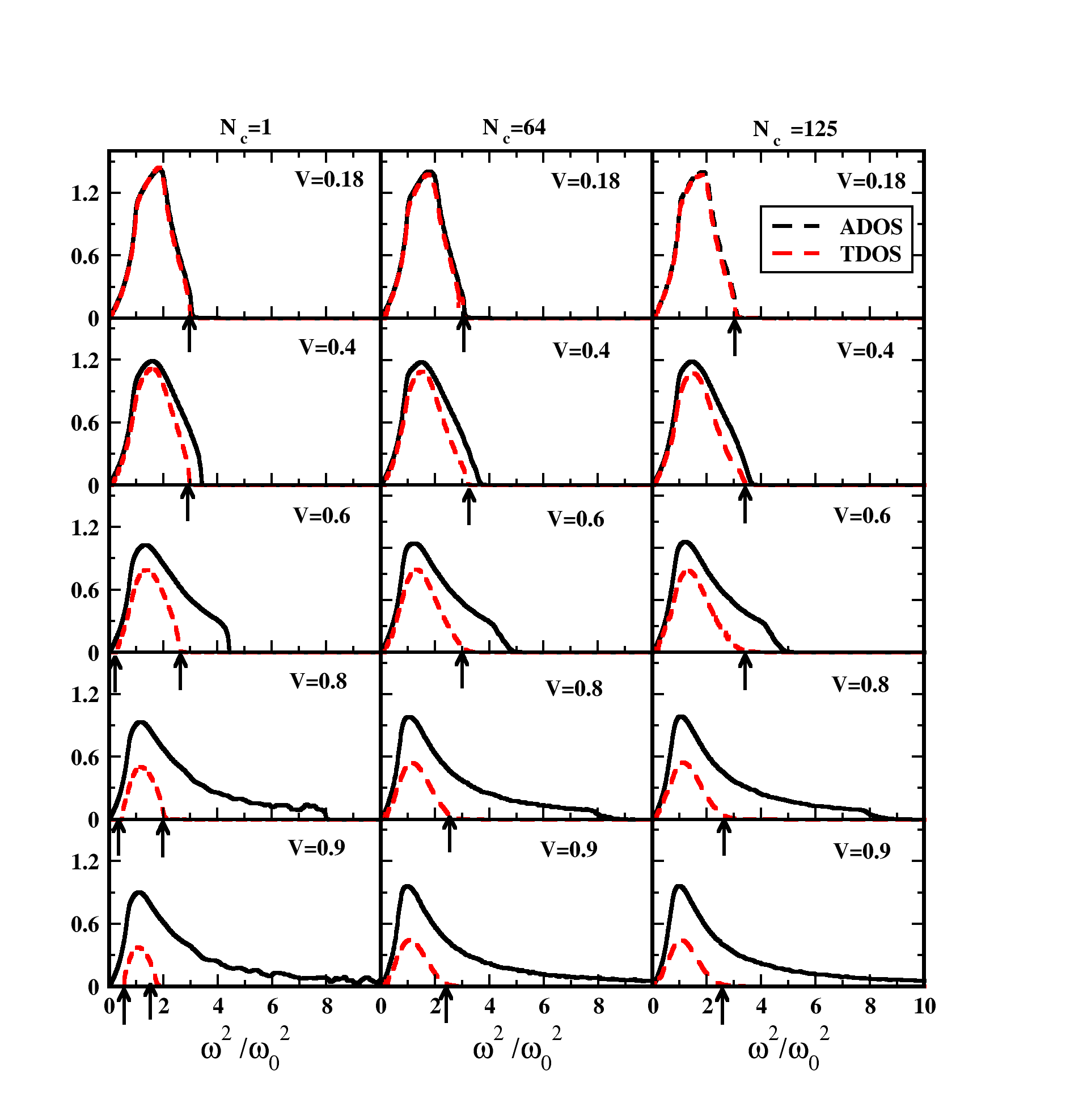}}
\caption{The evolution of the ADOS and TDOS, obtained from the TMDCA, as a function of the square of the frequency ($\omega^2$) at various disorder strengths $V$ chosen from a box distribution in three dimensions with cluster sizes $N_c =1, 64$ and $125$. At low disorder ($V$), the shape of the TDOS is similar to ADOS. As $V$ increases, the spectral weight in the TDOS decreases monotonically, while the ADOS, being normalized, develops
long, slowly decaying tails that comprise localized phonon modes. The tiny arrows denote mobility edges ($\omega_M)$, that have been used for benchmarking against TMM results in Fig.~\ref{fig:mobility_edge_phase}.}
\label{fig:TMDCA_evol_box}
\end{figure}

We restrict our discussion of TMDCA results to three-dimensional systems and focus first on box disorder. 
In Fig.~\ref{fig:TMDCA_evol_box}, the ADOS (black) and TDOS (red) are shown for a range of disorder strengths ($V$) and cluster sizes $N_c=1, 64$ and $125$ for a {\em uniform (box)} distribution. 
As may be expected, the typical DOS is almost the same as the ADOS for low disorder ($V\lesssim 0.4$). However, for higher $V$, localization sets in at higher frequencies. The ADOS develops long tails, but the TDOS is non-zero over a much smaller frequency interval, indicating that all tail modes are Anderson localized.  Moreover, the integrated spectral weight in the TDOS decreases steadily.  The TDOS shown in Fig.~\ref{fig:TMDCA_evol_box} has been used to extract the mobility edges that were compared against TMM results in Fig.~\ref{fig:mobility_edge_phase}. Note that, for higher $V (\gtrsim 0.8)$ and $N_c=1$,  the spectra exhibit a second mobility edge at low frequencies implying that long wavelength acoustic modes become localized. However, this result is again an artifact of the single-site approximation because higher $N_c$ results show that long wavelength acoustic modes do not localize at all, even when $V\rightarrow 1$.  

\subsection {Binary isotopic disorder}

 \begin{figure}[h!]
\centerline{\includegraphics[scale=0.7]{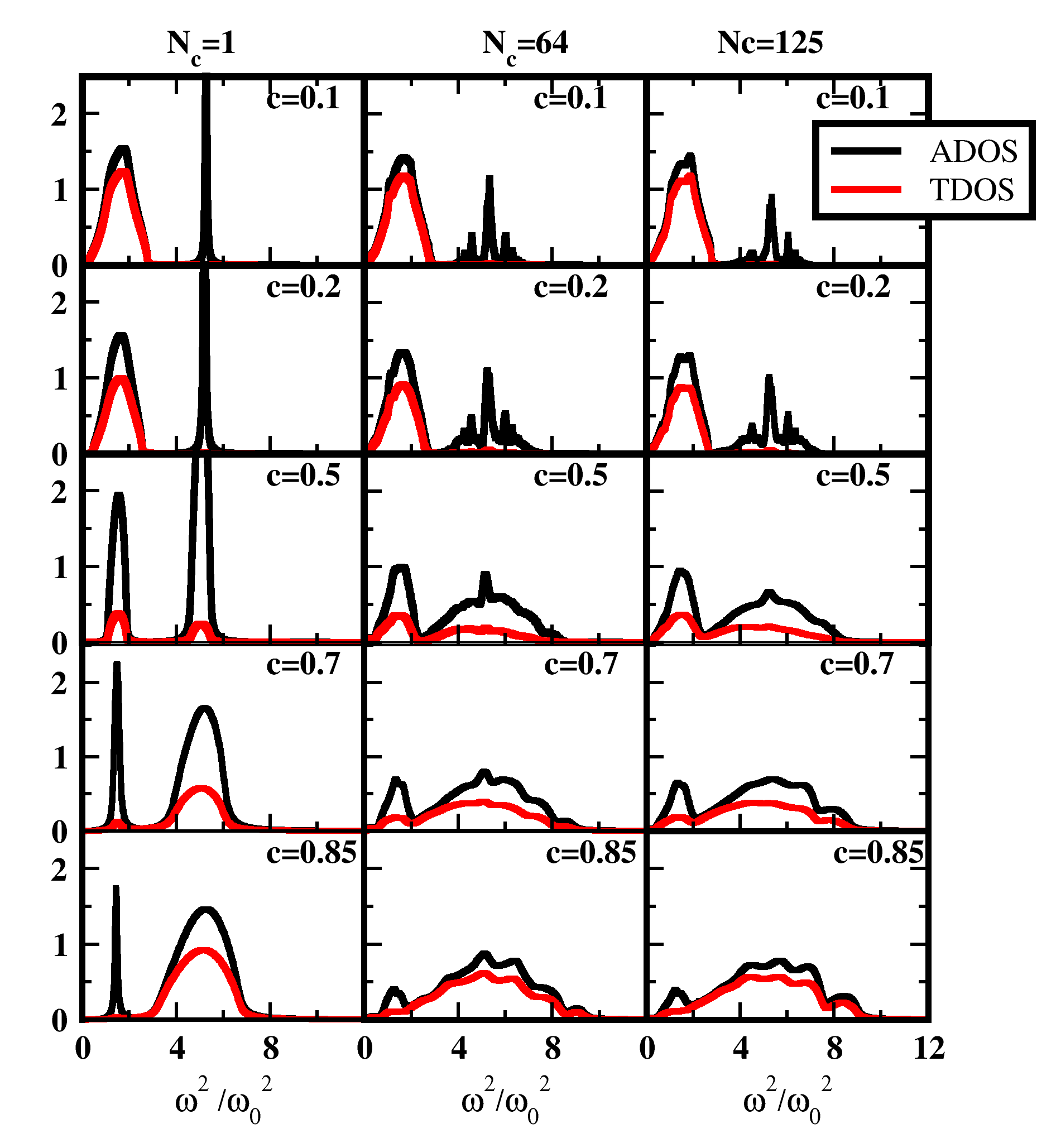}}
\caption{The ADOS and TDOS calculated using the TMDCA with cluster sizes $N_c = 1, 64$ and $125$ at various values of impurity concentration $c$ with fixed disorder potential $V=0.7$ for binary isotopic mass distribution in three dimensions.}
\label{fig:fig7_3d_binary_tmdca_c}
\end{figure}
For a binary, isotopic distribution, the evolution of ADOS and TDOS, obtained within the TMDCA for cluster sizes $N_c=1$, $N_c=64$ and $N_c=125$, with
increasing impurity concentration $c$ and fixed disorder potential $V=0.7$, is shown in  Fig.~\ref{fig:fig7_3d_binary_tmdca_c}.  Fig.~\ref{fig:fig7_3d_binary_tmdca_c} displays a transfer of spectral weight from low to high frequencies, and a modest dip in the typical spectral weight around $c=0.5$.

We find that the ADOS shown in   Fig.~\ref{fig:fig7_3d_binary_tmdca_c}  is almost the same than the one found within the DCA (see Fig.~\ref{fig:compare_ED_3D}). The main difference is that the ADOS found within the TMDCA is very spiky as compared to the corresponding quantity in the DCA.  Interestingly, the {\em impurity} modes yield a non-zero ADOS beyond the band-edge of the host band, but the TDOS is almost zero for low concentrations ($c\lesssim 0.2$). The vanishing of the TDOS indicates the localization of the impurity-induced high-frequency modes for such concentrations.  As the concentration increases, the low and high-frequency bands merge, and the TDOS is non-zero over the entire bandwidth. Nevertheless, as the concentration $c\rightarrow 1$, the ADOS clearly shows a remnant of the host modes, but the TDOS is quite small in the same frequency range implying that most of those modes are localized. The leftmost panel, for $N_c=1$, shows that for $c\rightarrow 1$, the host modes are completely localized, and a low-frequency mobility edge emerges. However, results for larger cluster sizes of $N_c=64$ and $125$ show that such a result is an artifact of ignoring non-local dynamical correlations. 

\begin{figure}[t]
\centerline{\includegraphics[scale=0.8]{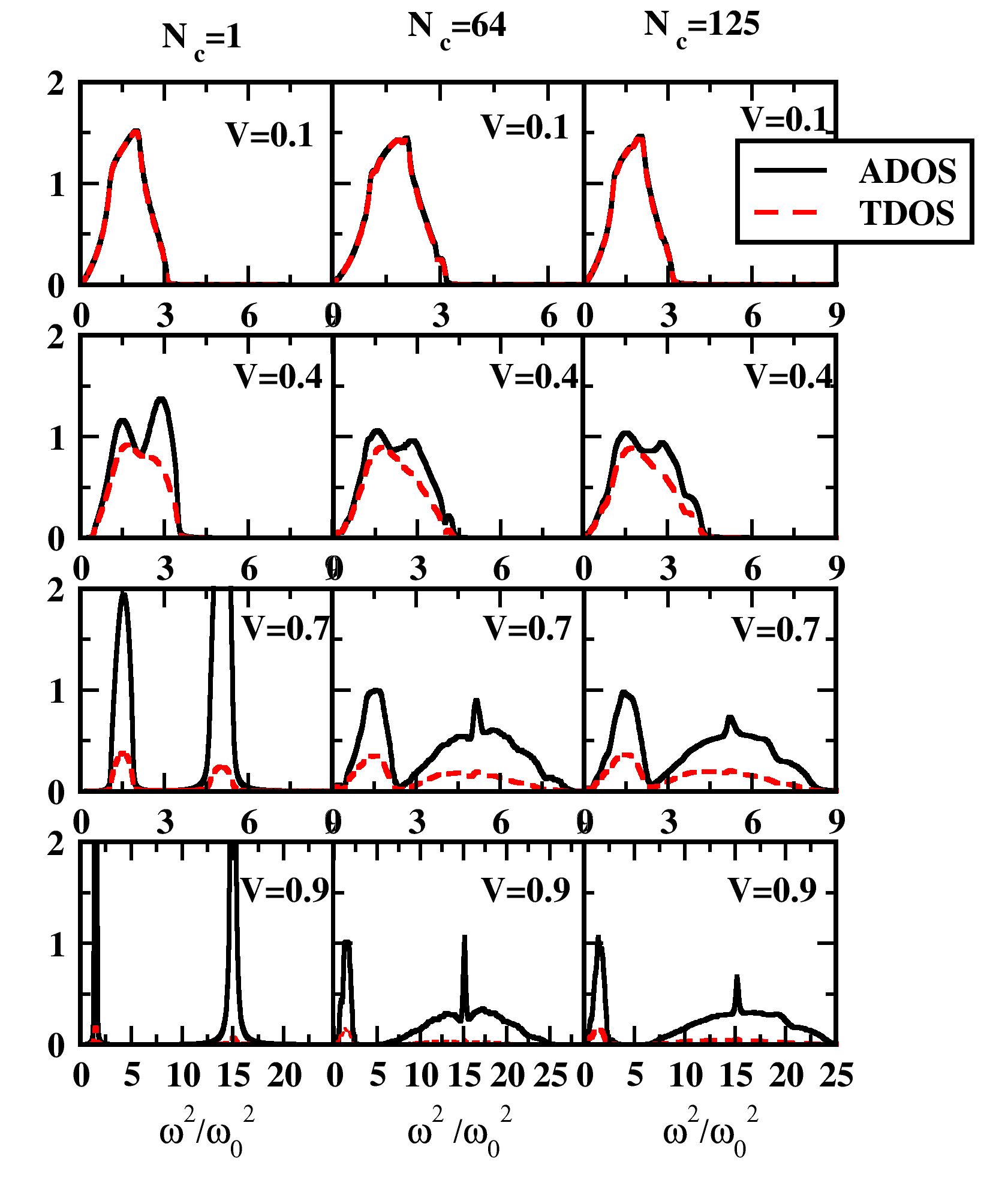}}
\caption{The ADOS and TDOS calculated using the TMDCA with cluster sizes $N_c = 1, 64$ and 125 at various values of disorder potential $V$ with impurity concentration $c=0.5$ for binary, isotopic mass distribution in three dimensions.}
\label{fig:fig8_3d_binary_tmdca_v}
\end{figure}
  Far more dramatic changes occur for fixed concentration, $c$,  and increasing disorder potential, $V$, as shown in Fig.~\ref{fig:fig8_3d_binary_tmdca_v}. At low $V$ ($\lesssim 0.4$), the ADOS and TDOS do not differ much, which is expected since the TMDCA reduces to DCA
 in the low disorder limit~\cite{PhysRevB.89.081107}. The good agreement between ADOS and TDOS also indicates that most modes remain propagating even if half of the host atoms are replaced with lighter atoms of mass, $M_{imp} \gtrsim 0.6 M_0$. However, for higher $V$, the TDOS is sharply suppressed and is seen almost to vanish for $V\rightarrow 1$, thus suggesting that almost all modes get localized in this parameter
regime.  Nevertheless, a complete localization seems to be possible only when $V=1$, or when $M_{imp}=0$, which corresponds to vacancies, for which a proper treatment involves the inclusion of spring disorder.

Therefore, localization of the impurity modes in the high-frequency region may be achieved
with experimentally feasible disorder parameters. However,
low-frequency phonons are almost impossible to localize,
which is consistent with the argument made in Section II.  
Howie et al. \cite {PhysRevLett.113.175501} study Hydrogen-Deuterium mixtures for three concentration ratios, namely $0.6:0.4, 0.55:0.45$, and $0.5:0.5$ using Raman spectroscopy. They observe that the Hydrogen-Deuterium mixture goes into a new phase IV, which may be modeled as an ideal binary isotopic alloy. In this alloy, with a mass-factor of 2 and varying the concentration ratio, they find six localized modes located in the high-frequency region,
while four low-frequency modes are found to be delocalized. 
Our model study using the TMDCA can capture this localization effect qualitatively. It will, naturally, be interesting to explore the phenomenon of acoustic phonon localization using more realistic parameters in the presence of both mass and spring disorder.
Such a study is underway.

The results shown in this section allow us to answer the second question posed at the end of the introduction. Although the ADOS
shows the physically observable exact spectrum of the disordered phonon system, a clear identification of localized and extended states cannot be made based only on the ADOS. Through a direct comparison of the TDOS with the ADOS, such an identification becomes straightforward. Thus, the TDOS gives great insight into which modes are propagative and 
which ones are not; that can be further used for developing strategies 
for e.g decreasing thermal conductivity in thermoelectric materials.

\section{Conclusions}
We have developed the DCA and TMDCA formalisms for investigating the effects of disorder on the phonon spectrum.  
Though the DCA exhibits several advantages over the CPA by including important non-local spatial correlations, it suffers from its inability to capture Anderson localization. Such a failure is due to  the arithmetic averaging over disorder configurations. Based on this understanding, we develop the TMDCA, where a typical averaging ansatz replaces the arithmetic averaging step. Using the TMDCA for a binary and a box distribution of mass disorder, we explore several aspects of Anderson localization in phononic systems. In particular, a comparison of the mobility edge computed through the TMDCA with that from the transfer matrix method yields an excellent agreement including the capture of the re-entrance transition of the mobility edge.

We also find that for a binary isotopic alloy, low concentrations of light impurities introduces high frequency modes, which are Anderson localized. While at high concentrations,
 the lower frequency modes are localized. Maximum localization over the entire spectrum is observed for equal concentrations of light and heavy atoms. Another finding is that a larger difference between the isotope masses introduces  stronger localization effects than the ones due to an increasing in the concentration of impurities. 
 
 Addressing the question 1(b) posed at the end of the introduction, the DCA and the TMDCA methods do fulfill several essential characteristics required for a successful cluster theory. They converge systematically to the thermodynamic limit, and with far lower computational expense than exact methods such as ED and TMM. The excellent benchmarks obtained show that not only do the methods work in the full parameter regime, and over all frequencies, the TMDCA is also capable of describing AL of phonons highly accurately. The equation for the Green's function, namely Eq.~\eqref{eq:eom}, is valid for mass and spring disorder, as well as for multiple branches. Thus, in principle, these methods should be able to go beyond mass disorder, and
 our preliminary results do support this conjecture. 
 Since these methods are computationally relatively inexpensive, it should be possible to incorporate material-specific information.  In combination with first principle approaches for phonons, the TMDCA can be an efficient tool for studying Anderson localization in real materials. However, for doing so, the present formalism should be extended to incorporate multiple non-degenerate branches and also to the inclusion of spring disorder in addition to mass disorder. Since the current formulation adopts the Green's function approach, it can be easily extended to layered geometries, thus allowing for investigations of phonon engineering in superlattice structures, heterostructures, thin films and interfaces. Some of these directions are presently in progress.

\begin{acknowledgments} 
A portion of this research (T.B.) was conducted at the Center for Nanophase Materials Sciences, which is a Department of Energy (DOE) Office of Science User Facility.  This material is also based upon work supported by the National Science Foundation under the NSF EPSCoR Cooperative Agreement No. EPS-1003897 with additional support from the Louisiana Board of Regents (M.J., W.R.M., J.M.). M. J acknowledges support from the DOE grant DE-SC0017861.
\end{acknowledgments}

\appendix

\section{Displacement-displacement Green's functions}
The Green's functions and their spectral representation for disordered lattice vibrations have been extensively discussed in the literature. However, for completeness, we re-derive some of those results here. 

The Hamiltonian for a mass disordered lattice within the 
harmonic approximation is written down as
\begin{equation}
H=\sum_{\alpha i l}\frac{{p^2_{i\alpha}(l)}}{2M_{i}(l)} + \frac {1}{2} \sum_{\alpha\beta ll^\prime ij} \Phi^{\alpha\beta}_{ij}(l,l^\prime)u_{\alpha}^i(l)u_{\beta}^j(l^\prime)\,,
\end{equation}
where the symbols and indices are described in Section II.
The retarded displacement-displacement Green's functions,
represented by,
\begin{equation}
iD^{ij}_{\alpha\beta}(l,l^\prime,t)=\langle\langle\, u^{i}_{\alpha}(l,t);u^{j}_{\beta}(l^\prime,0)\, \rangle\rangle\,, 
\label{eq:gfapp}
\end{equation}
may be found through the equation of motion formalism. Using the Heisenberg equation of motion, we get
\begin{multline}
\label{eqnofmotion1}
i \frac{\partial} {\partial t} \langle \langle u_{\alpha}^{i}(l,t);u_{\beta}^{j}(l^\prime,0) \rangle \rangle = i \delta(t) \langle \lbrack u_{\alpha}^{i}(l,t), u_{\beta}^{j}(l^\prime, 0)\rbrack \rangle\\
+ \langle \langle \lbrack u_{\alpha}^{i}(l,t),\mathcal H \rbrack ; u_{\beta}^{j}(l^\prime,0)\rangle \rangle. 
\end{multline}
Now, since $
\lbrack u_{\alpha}^{i}(l,t), \mathcal H \rbrack = i p_{i \alpha} (l)/M_{i}(l)
$, Eq.~\eqref{eqnofmotion1} can be written as
\begin{equation}
\label{eqnofmotion12}
\frac{\partial} {\partial t} \langle \langle u_{\alpha}^{i}(l,t); u_{\beta}^{j}(l^\prime,0) \rangle \rangle = 0 + \langle \langle \frac{p_{i\alpha} (l,t)}{M_{(i)}(l)}; u_{\beta}^{j}(l^\prime,0)\rangle \rangle.  
\end{equation}
A similar consideration for the momentum-displacement 
Green's function, $\langle \langle p_{i\alpha}(l,t) ;u_{\beta}^{j}(l^\prime,0) \rangle \rangle $, yields
\begin{multline}
\label{eqnofmotion2}
i \frac{\partial}{\partial t} \langle \langle \frac{p_{i\alpha} (l,t)}{M_{i}(l)};u_{\beta}^{j}(l^\prime,0)
\rangle \rangle
 = i \delta(t) \langle \left[ \frac{p_{i\alpha} (l)}{M_{i}(l)} , u_{\beta}^{j}(l^\prime)\right]\rangle \\
+ \langle \langle
 \left[ \frac{p_{i\alpha} (l,t)}{M_{i}(l)},\mathcal H  
 \right]; u_{\beta}^j(l^\prime,0) \rangle \rangle\,.
\end{multline}
Since
\begin{equation}
\left[ \frac{p_{i\alpha} (l,t)}{M_{i}(l)},\mathcal H  \right]=-i \sum_{\gamma,l^{\prime\prime} j^\prime} \frac{\Phi^{\gamma,\alpha}_{j^\prime,i}(l^{\prime\prime},l)}{M_{i}(l)} u_{\gamma}^{j^\prime}(l^{\prime\prime},t)\,,
\end{equation}
Eq.~\eqref{eqnofmotion2} reduces to
\begin{multline}
\label{eqnofmotion3}
i \frac{\partial}{\partial t} \langle\langle \frac{p_{i\alpha} (l,t)}{M_{i}(l)};u_{\beta}^{j}(l^\prime,0)\rangle\rangle = -i \frac{1}{M_i(l)} i \delta(t)\delta_{ij}\delta(l,l^\prime)\delta_{\alpha\beta} 
\\
-i \frac{1}{M_i(l)} \sum_{\gamma,l^{\prime\prime} j^\prime} {\Phi^{\gamma,\alpha}_{j^\prime i}(l^{\prime\prime},l)} \langle\langle u_{\gamma}^{j^\prime}(l^{\prime\prime}, t); u_{\beta}(l^\prime,0)\rangle\rangle\,.
\end{multline}
Taking derivative with respect to time on both sides of  Eq.~\eqref{eqnofmotion12} and using Eq.~\eqref{eqnofmotion3}, we get
\begin{multline}
\label{eqnofmotion4}
\frac{\partial^2} {\partial t^2} \langle\langle u_{\alpha}^{(i)}(l,t) u_{\beta}^{(j)}(l^\prime,0) \rangle \rangle = -\frac{1}{M_i(l)} i \delta(t) 
\delta_{ij} \delta_{\alpha\beta} \delta(l,l^\prime)\\
-\frac{1}{M_i(l)} \sum_{\gamma,l^{\prime\prime} j^\prime} {\Phi^{\gamma,\alpha}_{j^\prime i}(l^{\prime\prime},l)} 
\langle\langle
 u_{\gamma}^{j^\prime}(l^{\prime\prime}, t); u_{\beta}^{j}(l^\prime,0)\rangle \rangle \,. 
\end{multline}
Using the definition of Green's function (Eq.~\eqref{eq:gfapp}), we can re-write Eq.~\eqref{eqnofmotion4} as
\begin{multline}
\label{eqnofmotion5}
M_i(l)\frac{\partial^2} {\partial t^2} D_{\alpha \beta}^{ij}(l,l^\prime,t) = - \delta(t) \delta_{\alpha\beta}\delta_{ll^{\prime}}\delta_{ij}\\
-\sum_{\gamma,l^{\prime\prime} j^\prime} {\Phi^{\gamma \alpha}_{j^\prime i}(l^{\prime\prime}, l)} D_{\gamma \beta}^{j^\prime j}(l^{\prime\prime}, l^{\prime},t)\,.  
\end{multline}
Transforming to frequency space and using the symmetry relations of the force-constant matrix ($\Phi^{\gamma\alpha}_{j^\prime i}=\Phi^{\alpha\gamma}_{ij^\prime}$),  finally,  Eq.~\eqref{eqnofmotion5} can be written as
\begin{align}
M_i(l) \omega^2 D_{\alpha\beta}^{ij}(l,l^\prime,\omega) &= \delta_{\alpha\beta}\delta_{ll^\prime} \delta_{ij} \nnu \\
&+\sum_{\gamma,l^{\prime\prime}j^\prime} {\Phi^{\alpha \gamma}_{i j^\prime}(l, l^{\prime\prime})} D_{\gamma \beta}^{j^\prime j}(l^{\prime\prime}, l^\prime,\omega)\,. 
\label{eq:eomapp}
\end{align}

\section{Normalization condition in mass-disordered systems}
To obtain the normalization condition for the Green's function in the presence of mass-disorder, we expand the displacement (u) and momentum (p) in terms of normal modes as follows\cite{Maradudin1966273},
\begin{align}
u(l,t)&=\frac {1}{\sqrt{2M(l)}} \sum_s B^{(s)}(l) \sqrt{\frac{1}{\omega_s}}\times  \nnu \\
&\Bigg \lbrack b_s \exp(-i\omega_s t) + b_s^{\dagger} \exp (i \omega_s t) \Bigg \rbrack\\
p(l,t) &=\frac{1}{i} \sqrt{\frac{M(l)}{2}} \sum_s B^{(s)}(l) \sqrt{\omega_s}\times \nnu \\
& \Bigg \lbrack b_s \exp(-i\omega_s t) - b_s^{\dagger} \exp(i\omega_s t) \Bigg. \rbrack
\label{eq:norm1}
\end{align}
Here, $b_s$ and $b_s^{\dagger}$ are the phonon destruction and creation  operators for the $s^{\rm th}$ normal mode, respectively. Hence, they follows commutation algebra for bosons i.e $\left [b_s, b_{s^\prime}^{\dagger}\right]=\delta_{ss^\prime}$.
The normal modes $B^{s}(l)$ are defined by a quantum number s, which take $3p$ values for a three dimensional system with  $p$ ions in the basis. The normal modes satisfy orthonormality and completeness relations, namely
\begin{align}
\sum_l B^{(s)}(l)B^{(s^{\prime})}(l)&=\delta_{ss^{\prime}} \nonumber \\
\sum_s B^{(s)}(l)B^{(s)}(l^{\prime})&=\delta_{ll^{\prime}}.
\label{eq:norm2}
\end{align}
Inverting Eq.~\eqref{eq:norm1} to get 
the phonon creation ($b_s$) and annihilation operators ($b_{s}^\dagger$) in terms of displacement and momentum operators in frequency space, we get (using Eqs.~\ref{eq:norm2}),
\begin{align}
b_s &=\sum_l B^s(l)\frac{1}{\sqrt{2M(l) \omega_s}} \Big (M(l)\omega_s u(l,\omega)+i p(l,\omega)\Big)\\
b_s^\dagger& =\sum_l B^s(l) \frac{1}{\sqrt{2M(l) \omega_s}} \Big (M(l)\omega_s u(l,\omega)-i p(l,\omega)\Big).
\end{align}
Using the definition of displacement-displacement Green's function as given in Eq.~\eqref{eq:gfapp}, we get  
\begin{equation}
i D(l,l^\prime,\omega)=i  \frac{1}{\sqrt{M(l)M(l^\prime)}} \sum_s B^s(l)B^s(l^\prime)\frac{1}{{(\omega^+)}^2-\omega_s^2}.
\end{equation}
Thus, the normalization condition in mass-disordered systems is
\begin{equation}
-\frac{\rm Im}{\pi} \int^{\infty}_{0} d\omega\, (2\omega^+) \sqrt{M(l)}\,D(l,l^{\prime},\omega)\sqrt{M(l^\prime)}
=\delta_{ll^\prime}.
\end{equation}

\begin{figure}[b]
\includegraphics[scale=0.7]{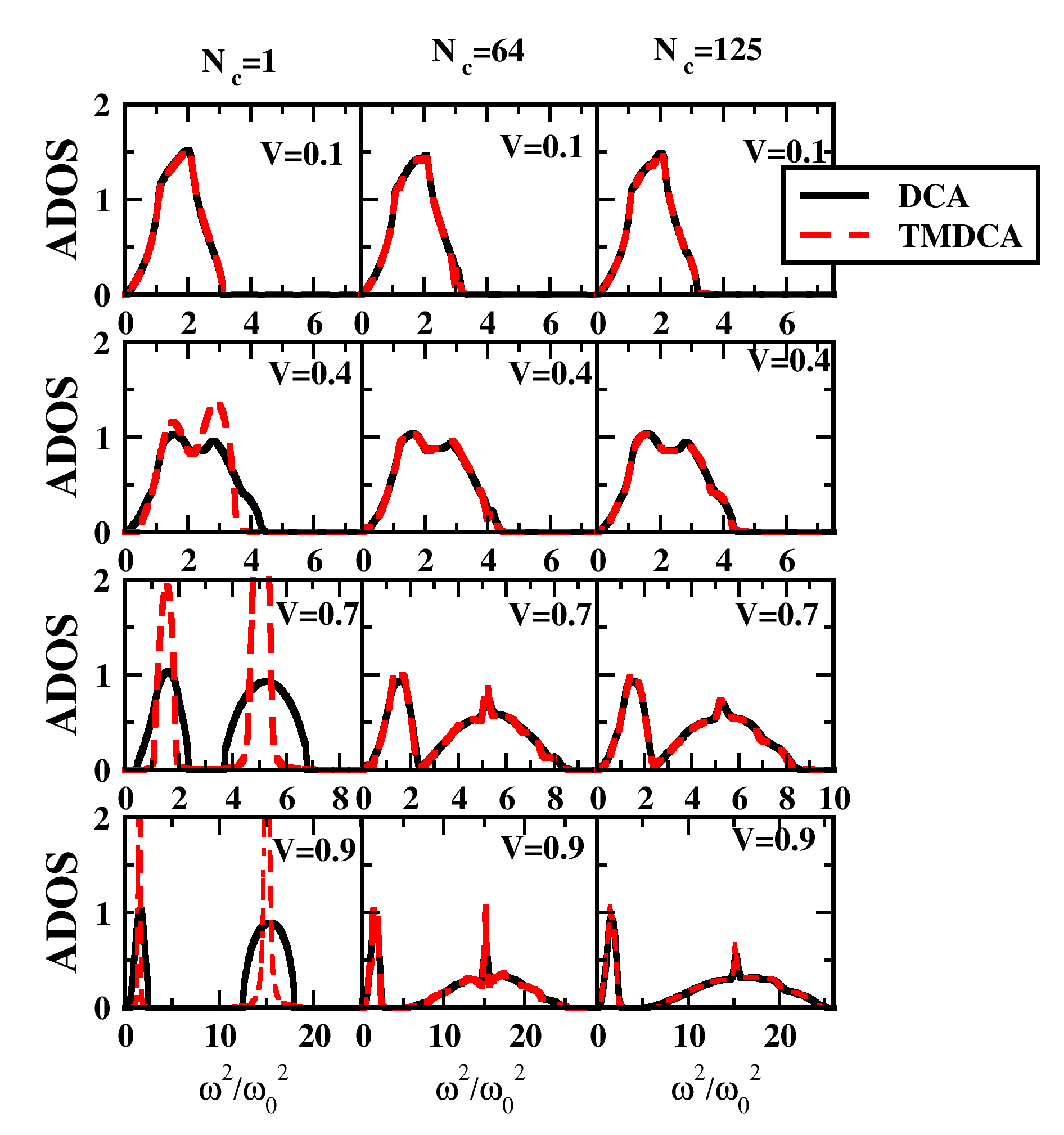}
\caption{The evolution of the ADOS calculated using the DCA (black curves) and TMDCA (red dashed curves) for cluster sizes of $N_c = 1, 64$ and $125$ at various values of disorder potential $V$ with fixed impurity concentration $c=0.5$ for a binary, isotopic distribution of masses in three dimensions. The ADOS obtained from the DCA and TMDCA differ significantly from each other for cluster size $N_c$=1, whereas for higher cluster size ($N_c$=64 and $N_c=125$), the two are completely identical to each other for all disorder potentials. This result indicates that at higher cluster size, ADOS is independent of hybridization function $\Gamma(\mathbf K, \omega)$, and equivalently the disorder averaging procedure. }
\label{fig:fig6}
\end{figure}
\section{Physical density of states from the DCA and the TMDCA}
In Fig.~\ref{fig:fig6}, we show results for the
arithmetically averaged phonon spectra computed within the DCA (black) and TMDCA (red) for a 
binary, isotopic mass distribution with fixed concentration $c=0.5$,
and various mass ratios ($M_{imp}/M_0$).

The main message here is that the physical density of states must not be dependent on the hybridization of the cluster provided that the cluster is large enough. And it is seen clearly in Fig.~\ref{fig:fig6} that the ADOS from DCA and TMDCA are identical for all disorder potentials for larger clusters, i,e $N_c=64$ and $125$. For $N_c=1$, the two differ
significantly at higher disorders, which is expected as mentioned above. However, ADOS is same for low disorder for all the cluster sizes $N_c=1, N_c=64$ and $N_c=125$, showing that TMDCA yields the same results as DCA at low disorder. Also, observe that the ADOS is the same for cluster sizes $N_c=64$ and $N_c=125$, which ensures the convergence of the results as cluster size increases.

\bibliographystyle{apsrev4-1}
\bibliography{apssamp}

\end{document}